\input harvmac
\input epsf
\epsfverbosetrue

\def\ap{\alpha'}

\def\b0{\bar{0}}
\def\b4{\bar{4}}
\Title{EFI-98-14}{\vbox{\centerline{'t Hooft Vortices and Phases of Large N 
Gauge Theory}
\vskip12pt
}}
\vskip20pt
\centerline{Miao Li}
\bigskip
\centerline{\it Enrico Fermi Institute}
\centerline{\it University of Chicago}
\centerline{\it 5640 Ellis Avenue, Chicago, IL 60637, USA}
\centerline{\it and}
\centerline{\it Institute for Theoretical Physics}
\centerline{\it University of California}
\centerline{\it Santa Barbara, CA 93106, USA}

\bigskip
\centerline{\it }
\centerline{\it }
\centerline{\it }
\bigskip

It is shown that a pair of vortex and anti-vortex is completely screened
in $2+1$ dimensional Yang-Mills theory and $3+1$ dimensional Yang-Mills 
theory in the strong coupling limit, based on the recent conjecture of 
Maldacena. This is consistent with the fact that these theories exhibit 
confinement.

\Date{April 1998}

\nref\th{G. 't Hooft, Nucl. Phys. B138 (1978) 1.}
\nref\cm{C. Callan and J. Maldacena, hep-th/9708147.}
\nref\mli{M. Li, hep-th/9803252.}
\nref\jm{J. Maldacena, hep-th/9711200.}
\nref\gkp{S. Gubser, I. Klebanov and A. Polyakov, hep-th/9802109.}
\nref\ewde{E. Witten, hep-th/9802150. }
\nref\coll{I. R. Klebanov, hep-th/9702076; S. Gubser, I. R. Klebanov
and A. A. Tseytlin, hep-th/9703040; M. R. Douglas, J. Polchinski
and A. Strominger, hep-th/9703031; A. Polyakov, hep-th/9711002.}
\nref\jmald{J. Maldacena, hep-th/9803002.}
\nref\ry{S. Rey and J. Yee, hep-th/9803001.}
\nref\minahan{J. Minahan, hep-th/9803111.}
\nref\witten{E. Witten,  hep-th/9803131.}
\nref\bitsy{A. Brandhuber, N. Itzhaki. Brandhuber, N. Itzhaki, 
J. Sonnenschein and S. Yankielowicz, hep-th/9803263.}
\nref\bisy{A. Brandhuber, N. Itzhaki, J. Sonnenschein and 
S. Yankielowicz, hep-th/9803137.}
\nref\rty{S. Rey, S. Theisen and J. Yee, hep-th/9803135.}
\nref\tamiaki{T. Yoneya, Nucl. Phys. B144 (1978) 195.}
\nref\imsy{N. Itzhaki, J. Maldacena, J. Sonnenschein and
S. Yankielowicz, hep-th/9802042.}

't Hooft suggested a disorder variable, dual to the Wilson loop, for
a non-Abelian gauge theory \th. The exchange algebra satisfied by this
operator and the Wilson loop operator enables a powerful argument for
possible phases of the gauge theory. Generalizing a solution to D-brane
Nambu-Goto action of \cm, it is proposed in \mli\ that a heavy 't Hooft
vortex on D2-branes corresponds to the point where an orthogonal
D2-brane intersects. A vortex line on a stack of D3-branes is T-dual
of this intersection point, and so on.

The recent proposal of duality between a large N strongly coupled Super
Yang-Mills theory and anti-de Sitter supergravity provides a powerful
tool to make many predictions for SYM \refs{\jm, \gkp, \ewde, \coll}.
In particular, the expectation value of a temporal Wilson loop can be
easily computed \refs{\jmald, \ry, \minahan, \witten}. The spatial
Wilson loop at a finite temperature exhibits the area law, indicating
confinement of the lower dimensional Yang-Mills theory without 
supersymmetry, thereby fulfills one of the old prophecies \refs{\witten,
\bitsy}. In addition, quark and anti-quark were shown to be screened
at a finite temperature if the separation is sufficiently large 
\refs{\bisy,\rty}.

In a pure Yang-Mills theory in $2+1$ dimensions or $3+1$ dimensions,
test quarks are confined, or there is a linear potential between a 
quark and an anti-quark. 't Hooft vortices or vortex-loops should
exhibit a different behavior. It is shown in \mli\ that in a maximally
supersymmetric Yang-Mills theory, the interaction potential between
a vortex and an anti-vortex obeys a power law. We therefore expect
this potential be weaker in a confining theory, such as a pure Yang-Mills
theory. The purpose of this short note is to show that indeed a complete
screening occurs in a confining theory, thereby to fulfill another old
prophecy. The total Green's function of two vortices is equal to the
disconnected part, which is given by the masses of vortices and 
thus displays the perimeter law in a trivial fashion.

To obtain a strongly coupled pure Yang-Mills theory in $p$ dimensional
spacetime, following Witten \witten, we start with Dp-branes and wrap the
Euclidean time around a circle of circumference $\beta=1/T$. Assigning
anti-periodic boundary condition to fermions gives mass of order $T$
to fermions. Scalars receive radiative correction to their mass at
one-loop, and hopefully the correction persists in the strong coupling
limit. Thus, at energies much lower than $T$, the theory is that of
effective $p$ dimensional pure Yang-Mills theory with Euclidean spacetime.
However, it is difficult to recover the weak coupling regime, due to
the singularity in the near horizon geometry caused by compactification
on a circle. For instance, it is not known how to directly see the
dimensional transmutation in the 4 dimensional Yang-Mills theory, namely
QCD, the most interesting case.

Consider D3-branes first. With a finite temperature, the classical
solution in the near horizon region is that of the anti-de Sitter black hole
\eqn\mt{ds^2/\ap ={U^2\over R^2}(f(U)dt^2+dX_i^2)+{R^2\over U^2}(f^{-1}(U)
dU^2+U^2 d\Omega_5),}
where $i=1,2,3$, $f(U)=1-(U_T/U)^4$. The dilaton is constant, and 
$R^2=(4\pi gN)^{1/2}$. The 4D Yang-Mills coupling $g_4^2\sim g$. We have 
Wick-rotated time. For the above geometry to be non-singular at the
horizon $U=U_T$, the circumference of $t$ must be $\beta =\pi R^2/U_T$,
or $U_T=\pi R^2 T$. 

We are interested in the effective 3D Yang-Mills theory, whose coupling
constant is given by $g_3^2=g_4^2T$. This sets a mass scale much smaller
than $T$. However, the physical mass scale is set by $g_3^2N$ which is
much greater than $T$. In order to avoid complicating theory with KK
modes, we should consider physics at distances much larger than $1/T$.
We shall take $X_1$ as the Euclidean time of the 3D world.

An orthogonal D3-brane intersecting the source D3-branes along $(t,X_1)$
creates a vortex in 3 dimensions. After dimensional reduction in $t$,
the intersection is parametrized by $X_1$, and is regarded as the world-line
of the vortex. When $N$ is finite, the solution clearly indicates that
this is a 't Hooft vortex \mli. The D3-brane can be regarded as a D2-brane
after dimensional reduction along $t$, and has a tension $T_3\beta$. Of course
this agrees with the D2-brane tension if one performs T-duality along $t$.
We shall set $\ap =1$ in the rest of the paper. $T_3$ measured in string
unit is a pure number $c_3=1/[(2\pi)^3g]$. We parametrize the other two
spatial dimensions of the orthogonal D3-brane by $(U, \theta)$, where
$\theta$ is embedded into the five sphere $S^5$. The total energy of a 
single D3-brane, taking $X_1$ as time, is given by
\eqn\dsing{E_0=c_3\beta\int d\theta\int_{U_T}^\infty UdU,}
where we assume that the D3-brane stretches all the way from spatial
infinity to the horizon. Notice that, although the background is not
supersymmetric, and the orthogonal D3-brane is not a BPS state, all the
nontrivial $U$ dependent factors cancel and the energy is equal to
that of a D3-brane embedded into a flat spacetime. The only difference
is that it does not extend to the origin $U=0$, which is the singularity
of metric \mt.

Before we set out to do further calculation, we want to pose and clarify
a puzzle. The existence of the 't Hooft vortex proposed in \mli\ 
depends crucially on the existence of a complex Higgs field. The configuration,
for a given $N$, is given by $(A, w)$, where $w$ is a complex $N\times N$
Higgs field which is singular at the location of the vortex. $A$ is a gauge
field having a nonvanishing holonomy around the vortex. Now, as we 
said before, in 3 dimensions the theory is effectively a pure Yang-Mills
theory, so
why does this vortex still exist? From the perspective of the 4D theory on
D3-branes, the answer is that classically the Higgs fields are still massless,
even though they gain a mass after turning on a temperature, the vortex
is so heavy that it still exists. By the same token, it also exists in the
3D gauge theory as a heavy probe. The same question arises when we use
the spatial Wilson loop to probe the 3D gauge theory. As shown in \cm,
the solution representing an open string attached to D-branes also depends
on the Higgs field crucially. Although there is no massless Higgs field
in the effective 3D theory, as a heavy probe the open string still
exists. As argued by Witten \witten, the properties of these probes
depend on the gauge field configuration, when the Higgs fields are given 
a mass. Also, the exchange algebra generated by the Wilson loop and
the 't Hooft vortex depends on the gauge field only \th.

Consider a pair of orthogonal D3-brane and anti-D3-brane. This system should
be regarded as a single D3-brane. The simplest topology is that given
in fig.1, where a throat connects the two D3-branes. When $T=0$, as shown
in \mli, there is always a saddle point of the Nambu-Born-Infeld action
with this topology. The energy of this saddle point is smaller than
that of two disconnected branes, thus it is more energetically favored.
The difference in energy is then interpreted as the interaction energy.
For $T>0$, no matter how small $T$ is, we shall show that for a separation
$L$ large than a certain value $L_{max}$, which is order $\beta$, there 
does not exist a saddle point with the topology in fig.1. There is another
scale $L_c<L_{max}$, also of order $\beta$, at which the energy of this
configuration is equal to that of the bare energy, and in between the
two points, the ``interaction energy'' is positive, therefore the 
connected configuration is not favored. We conclude that for a separation
$L>L_c$, the interaction energy vanishes. Interpreted in terms of
the 3D gauge theory, the vortex and the anti-vortex are completely
screened. As we shall explain later, this answer can not be the exact
one.

\bigskip
{\vbox{{\epsfxsize=2in
        \nobreak
    \centerline{\epsfbox{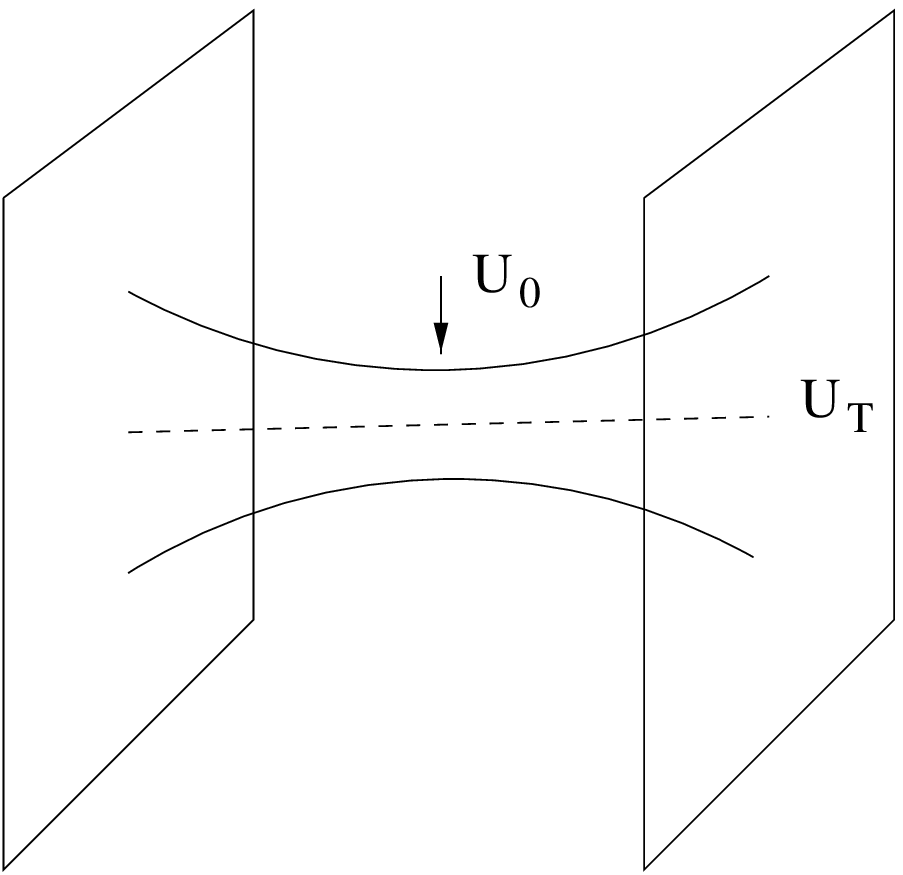}}
        \nobreak\bigskip
    {\raggedright\it \vbox{
{\bf Figure 1.}
{\it A D3-brane and an anti-D3-brane connected by a throat. In the Euclidean
metric, $U=U_T$ is the origin.}
}}}}}
\bigskip

Let the two D3-branes be separated in direction $X_2$ by a distance $L$,
for simplicity use $x$ to denote $X_2$. The configuration in fig.1 is
symmetric with respect to $x=0$, thus $U$ is an even function of $x$.
The induced world-volume metric on the D3-brane is
\eqn\imt{\eqalign{h_{tt}&={U^2\over R^2}f(U), \quad h_{11}={U^2\over R^2},\cr
h_{\theta\theta}&=R^2, \quad h_{xx}={U^2\over R^2}+{R^2\over U^2}f^{-1}
(U)(U_{,x})^2,}}
and the energy is given by
\eqn\dten{E=4\pi\beta c_3\int_0^{L/2}dx \left({U^6\over R^4}f(U)
+U^2(U_{,x})^2\right)^{{1\over2}}.}

The classical solution minimizing \dten\
\eqn\dts{x={R^2\over U_0}\int_1^{U/U_0}dy(y^4-\lambda)^{-{1\over 2}}
({y^2(y^4-\lambda)\over 1-\lambda}-1)^{-{1\over 2}},}
where $\lambda =(U_T/U_0)^4$. 
Given a value for $\lambda$, $L$ is determined by
\eqn\ddis{L={2\beta\over \pi}\lambda^{{1\over 4}}(1-\lambda )^{{1\over 2}}
\int_1^\infty dy(y^4-\lambda)^{-{1\over 2}}(y^6-\lambda y^2-1+\lambda
)^{-{1\over 2}},}
where we used the relation $U_T=\pi R^2 T$.

The interaction energy is obtained by subtracting from \dten\ twice 
of \dsing\
\eqn\pt{V=4\pi^3c_3R^4T\lambda^{-{1\over 2}}\left(\int_1^\infty dy
[y^2(y^4-\lambda)^{{1\over 2}}(y^6-\lambda y^2-1+\lambda )^{-{1\over
2}}-y]-{1\over 2}(1-\sqrt{\lambda})\right).}
For a sufficiently small $L$, the interaction approaches that in the
zero temperature case, except that here the interpretation is in
3D gauge theory rather than in the 4D SYM. $V$ is independent of
$g$. The small $L$ behavior is
\eqn\spt{V=-NTL^{-2}(c_1+c_2(LT)^4+\dots ),}
both $c_1$ and $c_2$ are positive constant. This is to be contrasted
to the zero temperature interaction energy in the 3D SYM, where the
power law is $L^{-4/3}$. Of course with small $L$ we are probing
the 4D SYM at finite temperature, not the 3D pure gauge theory.

It is seen from \ddis\ that not for all $L$ there is a solution. A
upper bound for the integral in \ddis\ is $\lambda^{-1/2}
\ln [(1+\sqrt{\lambda})/\sqrt{1-\lambda }]$. Combined with the
additional factor $\lambda^{{1\over 4}}(1-\lambda )^{{1\over 2}}$
it has a finite maximal value. Denote the maximal value of $L$ by
$L_{max}$ and the corresponding $\lambda$ by $\lambda_{max}$. A
numerical calculation shows $L_{max}=0.22 \beta$, $\lambda_{max}
=0.62$. Thus, at least beyond $L_{max}$ there is no saddle point solution
of topology in fig.1. We conclude that the interaction energy vanishes,
and vortices are completely screened.

However, the interaction rises to $0$ at the value $\lambda_c=
0.28$ (our boundary condition is $\lambda=0$, $L=0$), and the
corresponding separation is $L_c=0.196 \beta <L_{max}$. Beyond $L_c$,
the interaction energy becomes positive and one must discard
the saddle point. We then see that for $L>L_c$, the vortices
are screened. This value is comparable to the compactification
scale $\beta$. Indeed for this separation the theory is not effectively
3 dimensional yet.

In the above calculation, the only constraint on various parameters
is $g_4^2N\gg 1$. The three dimensional effective Yang-Mills coupling
is $g_3^2N=g_4^2NT$. We are exploring the infrared regime of the
3 dimensional theory if the separation between the vortex and the 
anti-vortex satisfies $g_4^2N(LT)\gg 1$. The critical value $L_c$
certainly satisfies this condition. Indeed, there is a range of $L$
smaller than $L_c$ also satisfies this condition. But since now $L$
is smaller than the compactification scale, we are no longer 
probing the 3 dimensional theory.

In the infrared regime, the 3 dimensional pure Yang-Mills theory is
believed to be confining. The test vortex is dual to a test quark in
the sense of \th, thus we expect screening. Our calculation however
does not imply that the interaction is exactly zero. In the maximal
SYM, the interaction falls off in $L$ as a power. The vortex
is coupled to an operator of color singlet, this power law can be
accounted for in principle. In a confining theory such as the pure
Yang-Mills, one expects that the exchange of color singlet states contributes
to the interaction, thus the interaction should fall off exponentially
with $L$, as there is a mass gap. As pointed out by Gross, such behavior
can rise in a quantum calculation where a saddle point does not exist.
Although the screening is perfectly understandable in a confining phase,
it is also interesting to ask precisely what screens the vortices,
causing the phenomenon similar to Debye screening, in a pure gauge theory.
It is difficult to construct a microscopic vortex state in
the continuum formulation, however, it is possible to do so on a lattice 
\tamiaki. Perhaps it is not far off the track to speculate that
the confining vacuum can be regarded as a medium of virtual vortices.

We proceed to the case of D4-branes. We obtain the 4D pure Yang-Mills theory
by compactifying the Euclidean time. This is the most interesting case,
since here we are making contact with QCD, although what we can say is
about the strong coupling phase. D4-branes can be obtained by wrapping
M5-branes around a circle, so in a sense the theory can be derived from
the conformally invariant one on M5-branes. There are two scales breaking
conformal invariance. The first circle, the M-circle, introduces a 
spacetime dependent dilaton. The second circle, the Euclidean time,
introduces a scale below which a pure Yang-Mills theory is obtained.

The near horizon geometry of D4-branes at zero temperature was discussed
in \imsy, and the finite temperature case was considered in \refs{
\witten, \bitsy}. The metric and dilaton with $\ap =1$ read
\eqn\dfmt{\eqalign{ds^2&={U^{{3\over 2}}\over a}(f(U)dt^2 +dX_i^2)
+aU^{-{3\over 2}}(f^{-1}(U)dU^2+U^2d\Omega_4^2),\cr
e^\phi &={a^{3/2}U^{3/4}\over 2\pi^{3/2}} N^{-1},}}
where $i=1,\dots , 4$, $f(U)=1-(U_T/U)^3$ and $a=g_{YM}\sqrt{N}/
(2\sqrt{\pi})$. $g^2_{YM}$ is the Yang-Mills coupling on D4-branes.
Written in the above form, the metric is independent of $N$, and the string
coupling scales with $N$ as $N^{-1}$, clearly indicating that this
is an attractive framework for dealing with QCD strings.
The relation between the circumference of $t$ and $U_T$ is
$\sqrt{U_T}=4\pi a/(3\beta)$.

To trust the classical supergravity, the value of $U$ must lie
in between $1/(g_{YM}^2N)\ll U\ll N^{1/3}/g_{YM}^2$ \bitsy. 
$U_T$ has to satisfy the same conditions. Since
the effective 4D gauge theory coupling is given by $g_4^2=g_{YM}^2T$,
this puts constraints on the 4D coupling: $1\ll g_4^2N\ll N^{2/3}$.
Thus the 4D pure gauge theory is in the strong coupling regime. 
In particular, the geometry \dfmt\ can not be used to describe the
asymptotic region, where the importance of the dimensional transmutation 
begins to appear.

An orthogonal D4-brane intersects source branes along $(t, X_1, X_2)$.
After dimensional reduction along $t$, and taking $X_1$ as the Euclidean 
time in the effective 4D theory, $X_2$ describes a vortex line, or
a 't Hooft loop. For a D4-brane stretching all the way to the horizon,
the bare energy per unit length along $X_2$ is
\eqn\dfen{E_0={2\pi^{3/2}N\beta c_4\over a^2}\int d\theta\int_{U_T}^\infty
dU,}
where $c_4=1/(2\pi)^4$. Note that the factor $N$ cancels with the one 
implicit in $a^2$.

A pair of D4-brane and anti-D4-brane again should be regarded as a 
single D4-brane as in fig.1. Let the separation along $x=X_3$ be 
$L$. The total energy per unit length along $X_2$ of the configuration is
\eqn\dften{E={4\pi^{5/2}N\beta c_4\over a^2}\int dxU\left({U^3\over a^2}
f+(U_{,x})^2\right)^{1/2}.}
Again the solution minimizing the above energy satisfies an equation similar
to \dts, and the relation between $L$ and $U_0$  is
\eqn\sdet{L={3\beta\over 2\pi}\lambda^{{1\over 6}}(1-\lambda )^{{1\over 2}}
\int_1^\infty dy (y^3-\lambda )^{-{1\over 2}}
(y^5-\lambda y^2-1+\lambda )^{-{1\over 2}},}
where we used the relation between $T$ and $U_T$, and $\lambda =(U_T/U_0)^3$.

Similar to the case studied earlier, there is a finite maximum for $L$.
Its numerical value is $L_{max}=0.21\beta$, and the corresponding
$\lambda_{max}=0.64$. Thus, we have a completely screening for $L>L_{max}$.

The interaction energy for a $L$ allowed by \sdet\ is 
\eqn\dfin{V=2^{-1}\pi^{-3/2}N\beta U_0^2a^{-2}
\left(\int_1^\infty dy[y^2(y^3-\lambda )^{{1\over 2}}
(y^5-\lambda y^2-1+\lambda )^{-{1\over 2}}-y]-{1\over 2}(1-\lambda^{2/3})
\right).}
For small separation, the interaction potential displays a power law
with an exponent different from the zero temperature case computed in
\mli. This is not surprising, since here the theory is effectively
5 dimensional rather than 4 dimensional.
The interaction remains negative for small $\lambda$ until it reaches
the value $\lambda_c=0.23$. This gives the separation, according to
\sdet, $L_c=0.184\beta$. This is smaller than $L_{max}$. For $L>L_c$,
the interaction energy becomes positive, and the saddle point solution
is not the favored configuration. Again both $L_{max}$ and $L_c$ are
comparable to $\beta$, and we can safely conclude that for a separation
much larger than the compactification scale, where the theory is effectively
4 dimensional, the vortex lines are screened.

The result of the fairly straightforward calculations in this note is
compatible with both the idea that an orthogonal D-brane
represents a heavy 't Hooft vortex, and the idea that the large N
gauge theory in the strong coupling limit can be described by classical
supergravity in the near horizon geometry. It would be very interesting
to study the behavior of vortices in a Higgs phase. To do this,
some matter fields are to be introduced, in order to higgs the vector
fields. We leave this problem for future.

\noindent{\bf Acknowledgments} 
We would like to thank T. Banks, J. de Boer, M. Douglas, J. Polchinski and
other participants to the duality program at Institute for Theoretical
Physics at University of California, Santa Barbara, for discussions.
We also thank D. Gross and the staff of ITP for warm hospitality.
This work was supported by DOE grant DE-FG02-90ER-40560 and NSF 
grant PHY 91-23780.

\listrefs
\end